\documentclass[
twocolumn,
 amsmath,amssymb,
 aps,
  twoside,
 superscriptaddress,
pra,
longbibliography
]{revtex4-1}

\usepackage{physics}
\usepackage{dsfont} 
\usepackage{color,,newfloat}
\usepackage{graphicx}
\usepackage{dcolumn}
\usepackage{bm}
\usepackage{hyperref}

\hypersetup{
    colorlinks=true,       
    linkcolor=blue,          
    citecolor=blue,        
    urlcolor=blue           
}

\usepackage{algcompatible}

\usepackage{etoolbox}

\AtBeginEnvironment{algorithm}{\vskip+15pt\noindent\par\nobreak\vskip-0pt}
\AtEndEnvironment{algorithm}{\noindent\hrulefill\par\nobreak\vskip+5pt}

\usepackage{newfloat}

\DeclareFloatingEnvironment[
    fileext=loa,
    listname=List of Algorithms,
    name=Algorithm,
    placement=tbhp,
]{algorithm}

\def\mytitle{Silicon photonic processor of two-qubit \\entangling quantum logic}
\begin{document}

\title{\vspace{-1.0cm}\LARGE\textbf{\textrm{\color{black} \mytitle}}}

\author{R. Santagati}
 \altaffiliation{These authors contributed equally to this work.}
 
 \altaffiliation{\\raffaele.santagati@bristol.ac.uk}
\author{J. W. Silverstone}
 \altaffiliation{These authors contributed equally to this work.}
  \altaffiliation{josh.silverstone@bristol.ac.uk}
\affiliation{
Quantum Engineering Technology Labs, H. H. Wills Physics Laboratory and Department of Electrical and Electronic Engineering, University of Bristol, BS8 1FD, UK. 
}
\author{M. J. Strain}
\author{M. Sorel}
\affiliation{School of Engineering, James Watt South Building, University of Glasgow, Glasgow G12 8QQ, UK }

\author{S. Miki}
\author{T. Yamashita}
\affiliation{National Institute of Information and Communications Technology, 588-2 Iwaoka, Kobe 651-2492, Japan}

\author{M. Fujiwara}
\author{M. Sasaki}
\affiliation{National Institute of Information and Communications Technology, 4-2-1 Nukui-Kitamachi, Koganei, Tokyo 184-8795, Japan}

\author{H. Terai}
\affiliation{ National Institute of Information and Communications Technology, 588-2 Iwaoka, Kobe 651-2492, Japan}

\author{M. G. Tanner}
\author{C. M. Natarajan}
\author{R. H. Hadfield}
\affiliation{School of Engineering, James Watt South Building, University of Glasgow, Glasgow G12 8QQ, UK }

\author{J. L. O'Brien}
\author{M. G. Thompson}
\affiliation{
Quantum Engineering Technology Labs, H. H. Wills Physics Laboratory and Department of Electrical and Electronic Engineering, University of Bristol, BS8 1FD, UK. 
}

\begin{abstract}

\textbf{Abstract:}
Entanglement is a fundamental property of quantum mechanics, and is a primary resource in quantum information systems. Its manipulation remains a central challenge in the development of quantum technology. In this work, we demonstrate a device which can generate, manipulate, and analyse two-qubit entangled states, using miniature and mass-manufacturable silicon photonics. By combining four photon-pair sources with a reconfigurable six-mode interferometer, embedding a switchable entangling gate, we generate two-qubit entangled states, manipulate their entanglement, and analyse them, all in the same silicon chip. Using quantum state tomography, we show how our source can produce a range of entangled and separable states, and how our switchable controlled-Z gate operates on them, entangling them or making them separable depending on its configuration.
\end{abstract}

\maketitle

\section{Introduction}

Photons remain a promising vehicle for the development of next-generation quantum technology~\cite{Obrien:2009un, Latmiral:2016dp}. Integrated quantum photonics, with its intrinsic phase stability and miniature devices, is necessary to bring linear optics to the large scale \cite{Politi2008, Metcalf:2014jwa, Minkov:2016fm}. Several integrated photonic platforms have emerged to solve this problem, 
including silica-on-silicon~\cite{Politi2008, Matsuda:2014cy, Reimer:2015bva, Carolan:2015fb}, direct-write glass~\cite{Sansoni2010, Tillmann:2013jva, Flamini:2015cb, Bentivegna:2015iaba,Spring:2017gj}, lithium niobate~\cite{Vergyris:2016bd, Alibart:2016jo, Lenzini2017, Sansoni:2017jt}, silicon nitride~\cite{Zhang:2016dy, Moss:2013kv} and silicon-on-insulator~\cite{silverstone2016}. Silicon quantum photonics promises to simultaneously achieve the required functionality, performance, and scale.

Several important quantum optical functionalities have already been shown with high performance in silicon. Photon pairs can be generated using spontaneous four-wave mixing (SFWM)~\cite{Sharping2006,  Azzini2012, Matsuda:2012dma, Olislager2013, Collins:2013eu,  Xiong:2016bv}, and interfered with high visibility~\cite{Harada:2011cw, Silverstone:2013fu,   Takesue:2014ic, Xiong:2016bv, Xu:2013jna}. Single-photon \cite{Silverstone2015} and pump-rejection \cite{Harris2014, Piekarek:dUPyT_rs} spectral demultiplexers, as well as two-mode interferometers \cite{Wilkes:2016ba}, have been demonstrated with very high extinction. Finally, single-photon detectors, based on superconducting nanowires have shown excellent performance on silicon waveguides~\cite{Najafi2015, Pernice:2012bc}. The very high refractive index contrast of silicon-on-insulator waveguides yields micron-scale components (e.g.~\cite{Xu:2008db}), while miniature ring resonator SFWM sources~\cite{Azzini2012}, and quantum interferometric networks~\cite{Harris:2014kz} facilitate devices on a very large scale.

The integration of entangled qubit sources with entangling quantum logic, together on a common platform, is an important next step. Here we show a new method for generating path-encoded, variably entangled two-qubit states. We perform multi-qubit quantum logic on these states and study their entanglement. We implemented this scheme on a reconfigurable, silicon photonic device to generate a wide range of two-qubit states. We integrated this source with arbitrary state preparation, a switchable two-qubit gate, and an interferometer for tomographic analysis. The implemented quantum circuit is similar to the one reported in~\cite{Shadbolt2012}.

We tested the device's quantum logic capabilities with several experiments. We analysed the source performance using reversed-Hong-Ou-Mandel-type (RHOM)~\cite{Chen:2007he, Silverstone:2013fu} quantum interference, and qubit tomography on a wide range of possible states. We followed this with an exploration of the on-chip quantum logic, with the switchable two-qubit gate in both entangling ($\hat{\textsc{cz}}$) and non-entangling ($\hat{I}$) configurations, and using the purity ($P$)~\cite{Gamel:2012hm}, the CHSH parameter ($S$)~\cite{Horodecki:2009gb} and the Schmidt number ($K$)~\cite{Terhal:2000gd} as diagnostic metrics.

\begin{figure*}[t!]
\centering
\includegraphics[width=1\linewidth]{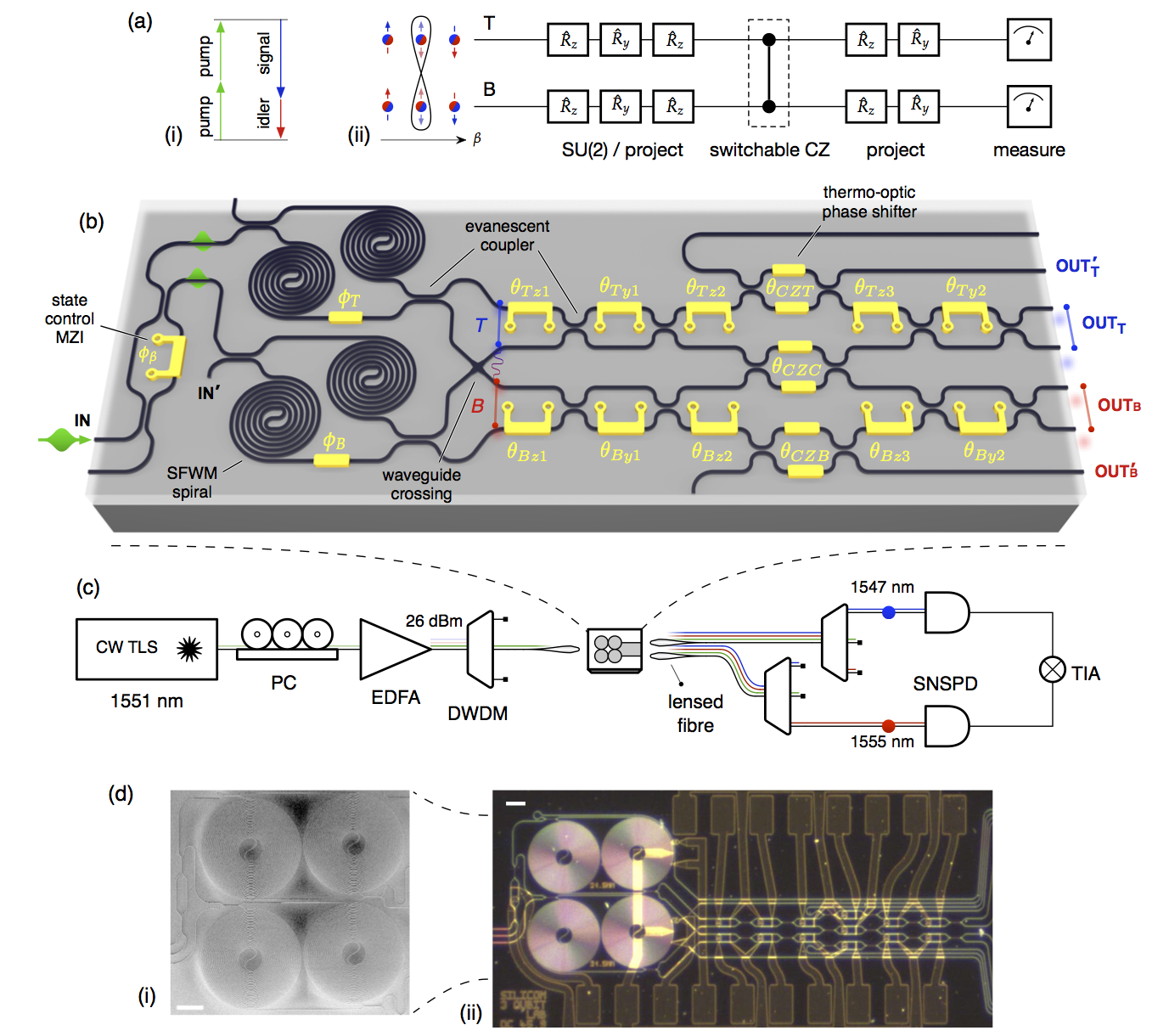}
\caption{Device and apparatus overview. \textbf{a}~Operating principles. \textbf{i}~Non-degenerate spontaneous four-wave mixing, \textbf{ii}~quantum circuit description. \textbf{b}~Schematic of the silicon quantum photonic chip. A pump laser is coupled into the device, coherently pumping two spiralled RHOM sources which produce two photons entangled or separable in path. These are fed into a reconfigurable linear optical network which can entangle or disentangle them, and analyse the output.  \textbf{c}~Off-chip apparatus. A continuous wave (CW) tunable laser source (TLS) is polarisation controlled (PC), amplified (EDFA), filtered and coupled onto the chip using lensed fibres and spot-size converters. Signal, idler, and pump photons coupled back into fibre in the same way, then spatially separated using dense wavelength-division multiplexers (DWDM), detected using superconducting nanowire single-photon detectors (SNSPD), and the output signal is analysed by a time interval analyser (TIA). \textbf{d}~Electron \textbf{i}~and optical \textbf{ii}~micrographs of the device.}
\label{figure1}
\end{figure*}

\section{Device structure and operation}

A schematic of the device is shown in Figure \ref{figure1}a. It comprises a reconfigurable source of two path-encoded entangled photons, controlled by the parameters $\phi_\beta$, $\phi_T$ and $\phi_B$. The source is followed by a reconfigurable interferometer, able to implement any two-qubit projector (including entangled projections). This second part of the device can be divided into three sections: arbitrary single qubit gates, a switchable post-selected controlled-Z ($\hat{\textsc{cz}}$gate) gate \cite{Ralph:2002id}, and final single-qubit unitaries, used to implement projectors for quantum state tomography, to reconstruct the output state.

The device comprised $500\times220\,\mathrm{nm}^2$ waveguides, directional couplers (approximate length  $45.9~\mu\textrm{m}$), a waveguide crossing ($> 20\ \mathrm{dB}$ isolation), and resistive metallic heaters (length $54.0~\mu\textrm{m}$). It was coupled to fibre via edge coupling, fibre lenses, and polymer spot-size converters. Electrical connections were achieved through multi-contact electrical probes and $200-\mu\textrm{m}$-pitch on-chip gold pads (approximately $120~\times~200~\mu\textrm{m}^2$). Fabrication of the device proceeded as in reference \cite{Silverstone2015}.
    
The experimental setup is presented in Figure \ref{figure1}b. Photons are generated on the chip via SFWM, pumped by an amplified continuous-wave tunable laser, and filtered to remove in-band noise. An average facet-to-facet transmission of $\approx -28\ \mathrm{dB}$ was observed. The dominant sources of loss were scattering at the chip facets, and propagation loss in the spiralled source waveguides. Inside the device the light was reconfigurably manipulated by an interferometric network, composed of evanescent coupler beam-splitters and thermo-optic phase-shifters~\cite{Trinh:1995gm, Harris:2014kz}. Photons were collected from the device, demultiplexed and separated from the pump using dense wavelength-division multiplexers (DWDM), detected using superconducting nanowire detectors~\cite{Miki:2013cv}, and finally converted into coincidence counts by a time-interval analyser.

\subsection{Photon-pair generation}

The strong non-linear properties of silicon waveguides are well known \cite{Leuthold:2010dg}. Spontaneous four-wave mixing (SFWM), an effect of the $\chi^{(3)}$  non-linearity, is now commonly used to produce photon pairs in silicon quantum photonic devices \cite{Sharping2006, Silverstone:2013fu}.

In the non-degenerate SFWM process used here, two photons from a bright pump are annihilated, producing two correlated photons with different wavelengths (Figure \ref{figure1}a). The two generated photons, `signal' and `idler', emerge spectrally on either side of the pump, conserving energy and momentum. In our experiment, spiralled 21-mm-long waveguides were used to produce photon-pairs, with the pump, signal, and idler photon wavelengths being 1551~nm, 1547~nm, and 1555~nm. These photons were generated in a continuous spectrum and the chosen wavelengths were post-selected by the off-chip demultiplexers.

\begin{figure}[tbh!]
\centering
\includegraphics[width=\linewidth]{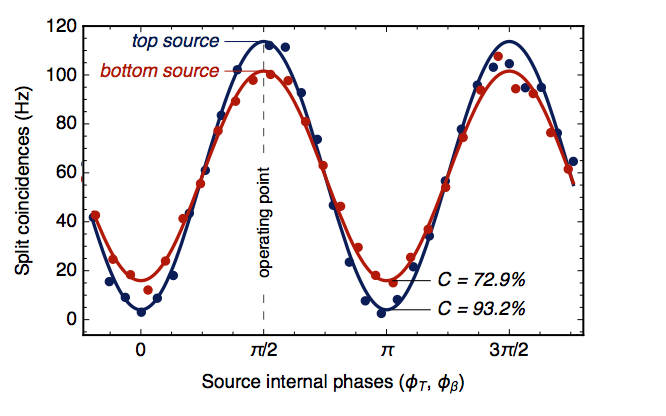}
\caption{Quantum interference for the two sources, measuring coincidences from the outputs $\mathrm{OUT}_T'$ and $\mathrm{OUT}_B'$, obtained by pumping each RHOM source and scanning the source internal phase, $\phi_T$ or $\phi_B$. The imperfect interference can be explained in terms of imbalance in the on-chip evanescent coupler beam splitters.}
\label{figure2}
\end{figure}

\begin{table*}[bt!]
\centering
\begin{tabular}{cc|cccc}
\textbf{Source state} & \textbf{Gate} & \textbf{Purity} $P$	& \textbf{Schmidt number} $K$  &	\textbf{CHSH} $S$ &	\textbf{Fidelity} $F'$\\
\hline
 \rule{0pt}{4mm}
 $|00\rangle$ & {bypassed} & $0.995 \pm 0.012 $  & $1.012 \pm 0.011$   & $1.577\pm 0.072$  & $0.973 \pm 0.011$ \\
 $|00\rangle$ & {$\hat{I}$} & $0.946 \pm 0.031 $ & $1.034 \pm 0.017 $ &  $1.465 \pm0.064 $ & $0.962 \pm 0.016$\\
 \rule{0pt}{5mm}
 $|11\rangle$ & {bypassed} & $0.998 \pm 0.008$ & $1.004 \pm 0.006 $ & $1.511 \pm 0.049$ & $0.984 \pm 0.007 $\\
 $|11\rangle$ & {$\hat{I}$} & $0.949 \pm 0.055 $ & $1.048\pm 0.037 $ & $1.601 \pm 0.121$ & $0.948 \pm 0.031$ \\
 \rule{0pt}{5mm}
 $(|00\rangle + |11\rangle)/\sqrt{2}$ & {bypassed} & $0.864 \pm0.019$ & $1.905 \pm 0.022 $ &  $2.560 \pm 0.037$  & $0.909 \pm 0.028 $\\
 $(|00\rangle + |11\rangle)/\sqrt{2}$ & {$\hat{I}$} & $0.832 \pm 0.040$ & $1.936\pm 0.025$  & $2.538\pm0.072$ & $0.900 \pm 0.026$\\
 \rule{0pt}{5mm}
 $|++\rangle$ & {$\hat{\textsc{cz}}$} & $0.931 \pm0.036$ & $1.657 \pm 0.045 $ & $2.560 \pm 0.078$ & $0.873 \pm 0.038$  \\
 $(|00\rangle + |11\rangle)/\sqrt{2}$ & {$\hat{\textsc{cz}}$} & $0.900 \pm 0.071$ & $1.166 \pm 0.055 $&  $1.907 \pm 0.137$ &  $0.839 \pm 0.013$
\end{tabular}
\caption{Purity, Schmidt number, CHSH parameters and Fidelity for a variety of measured states. The Schmidt number and CHSH parameter indicate entanglement. $S > 2$ indicates the presence of non-local correlations~\cite{Horodecki:2009gb}, while $K$ indicates the number of coefficients in the Schmidt decomposition of the state~\cite{Terhal:2000gd}. The fidelities $F'$ reported are computed against the ideal state optimised over local $R_z$ rotations, to compensate for the intrinsic random phase factor on each qubit.}\label{table:data}
\end{table*}

\subsection{Entangled qubit generation}

Our device uses a new scheme to generate entangled path-encoded states, which can subsequently be interfered, using pairs of \emph{non-degenerate} photons. Pump laser is distributed between two reverse-HOM structures using a reconfigurable power splitter (splitting ratio $\sin^2[\phi_\beta/2]$). Each RHOM contains two spiralled waveguides and a thermal phase shifter, as in \cite{Silverstone:2013fu}. The internal RHOM phases ($\phi_T$ and $\phi_B$) were set to $\pi/2$, such that the produced photon-pairs emerged deterministically split, one in each output waveguide, and in a state symmetrical between signal and idler photons. $\phi_\beta$ allows us to control the balance of photon-pair emission between the two RHOM structures, and so to control the entanglement present in the two-qubit output state.

Following Figure \ref{figure1}b, if $\phi_\beta = \pi$, photons will be generated only in the top RHOM, and the photon number output state, after the waveguide crossing, will be $|\mathit{1010}\rangle$, or $|00\rangle$ in the qubit basis. On the other hand, if $\phi_\beta = 0$, only the bottom RHOM generates photons, leading to $|\mathit{0101}\rangle = |11\rangle$.  Finally, if $\phi_\beta = \pi/2$, we obtain the maximally entangled state: $|\Phi^\Theta\rangle \equiv (|00\rangle+e^{i \Theta}|11\rangle)/\sqrt{2}$, where $\Theta$ is a fixed phase factor due to the chip's intrinsic path-length mismatch. Thus, the output state from the entangled qubit generator is
\begin{equation}
  |\psi\rangle = \sqrt{\beta}|00\rangle + e^{i \Theta} \sqrt{1-\beta}|11\rangle
\label{equation1}
\end{equation}
which can be continuously varied across a wide range of separable and entangled states, depending on the balance parameter, $\beta$. The balance depends on the square of the power division of the state control MZI (controlled by the phase $\phi_\beta$), due to the two-photon dependence of SFWM:
\begin{equation}
  \beta = \left|\frac{\sin^2(\phi_\beta/2)}{\sqrt{\sin^4(\phi_\beta/2) + \cos^4(\phi_\beta/2)}}\right|^2.
  \label{bal}
\end{equation}

\subsection{Quantum logic and analysis}

The state $|\psi\rangle$ is fed into a two-qubit circuit, composed of single-qubit rotations, and a switchable entangling gate. We implemented the arbitrary rotations on each qubit by cascading phase-shifters and Mach-Zehnder interferometers (MZI). These were used to realise  $\hat R_z$  and $\hat R_y$ rotations, respectively, obtaining an arbitrary $\mathrm{SU}(2)$ with the combination $\hat R_{z}\cdot \hat R_y \cdot \hat R_{z}$.

We implemented a switchable entangling gate using a scheme based on~\cite{Ralph:2002id}, but replacing the $1/3$ beam-splitters with tunable-reflectivity MZIs. In this way, we can switch the gate's controlled-Z operation on and off. When on, the $\hat{\textsc{cz}}$gate operation succeeds with probability $1/9$. 
In the remaining $8/9$ cases non-qubit states are generated, which are filtered by the coincidence-counting post-selection. 

Note that only the on ($\cos{(\theta_{CZ})} = 1/3$) 
and off ($\cos{(\theta_{CZ})} = -1$) 

gate configurations produce unitary operations. 
The two qubit gate is followed by rotations (parametrised by $\theta_{Mz3}$, $\theta_{My2}$, 
$M\in\{T,B\}$) used to implement quantum state tomography, via the method described in~\cite{James:2001bb}.

\subsection{Calibration}

Since the phase shifter parameters (phase-per-electrical-power, and phase offset) varied between phase modulators, a calibration process was essential. Measuring the bright-light transmission from the inputs ($\mathrm{IN}$ and $\mathrm{IN}'$) to the outputs ($\mathrm{OUT}_T$, $\mathrm{OUT}_B$, $\mathrm{OUT}_T'$, $\mathrm{OUT}_B'$),
we were able to characterise the electro-optic parameters of each thermal phase shifter, in a similar way to that described in \cite{Santagati2016}. We learned the parameters associated with each phase according to the scheme:
\begin{equation}
\begin{split}
\mathrm{IN}' \rightarrow \mathrm{OUT}_T', \mathrm{OUT}_B' &: \phi_{B}, \theta_{By1}, 
\theta_{\mathrm{CZB}}, \theta_{Ty1}, \theta_{\mathrm{CZT}}\\
\mathrm{IN} \rightarrow \mathrm{OUT}_T', \mathrm{OUT}_B' &: \phi_\beta, \phi_{T}, \theta_{Tz1}, \theta_{Bz1}\\
\mathrm{IN} \rightarrow \mathrm{OUT}_T &: \theta_{\mathrm{CZC}}, \theta_{Ty2}, \theta_{Tz2}, \theta_{Tz3}\\
\mathrm{IN} \rightarrow \mathrm{OUT}_B &: \theta_{By2}, \theta_{Bz2}, \theta_{Bz3}.
\end{split}
\label{eq:calibration}
\end{equation}
We observed instabilities in the calibration data, due to changes in electrical contact resistance between our probe card and the on-chip gold pads. To mitigate this, we periodically recalibrated the on-chip parameters. Metallurgical wire-bonded contacts can prevent this in future. Low levels of thermal and common-ground crosstalk were observed but not compensated. Recent results suggest that crosstalk can be reduced through efficiency improvements, passive compensation methods, and by current driving of the thermal phase shifters~\cite{Harris:2014kz, Santagati2016, Paesani:2017ga}.

The offsets of the tomographic $z$-rotation phases ($\theta_{Tz3}$, $\theta_{Bz3}$) were left at zero, meaning that additional random (fixed) $z$ rotations were applied to each qubit before measurement. This choice was necessitated by the combined difficulty of: (1) calibrating the non-linear source phase with bright light, and (2) doing this for each setting of the gate, in the device's finite stability time.

\section{Results}

\begin{figure}[tbh!]
\centering
\includegraphics[width=\linewidth]{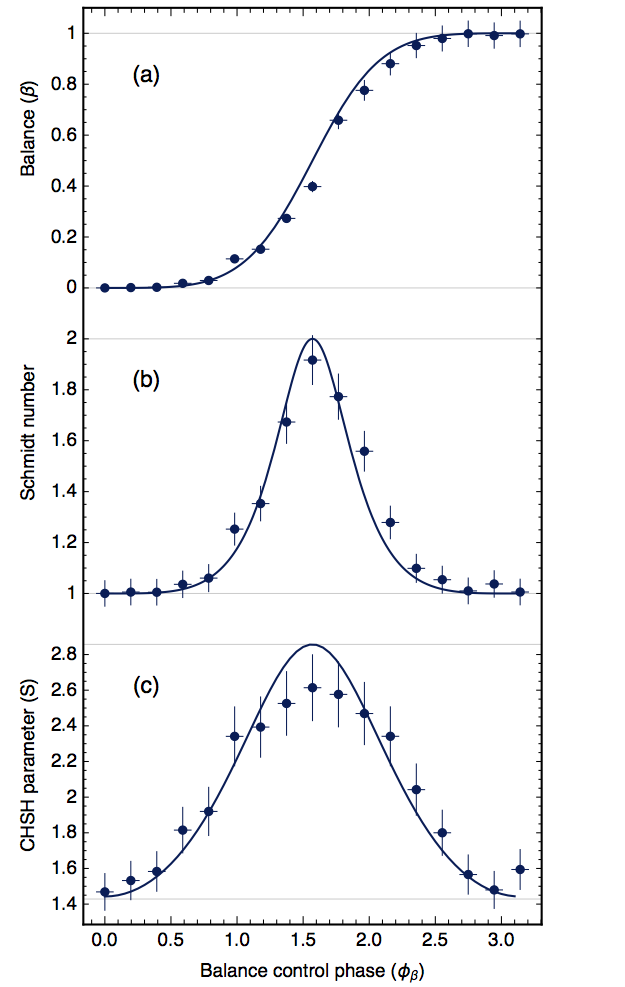}
\caption{Two-qubit state properties, direct from the source, as a function of the input state control phase, $\phi_\beta$. \textbf{a}~Balance between the $|00\rangle$ and the $|11\rangle$ components of the state, see equation \eqref{bal}. \textbf{b}~Schmidt number. \textbf{c}~CHSH parameter. Maximal entanglement occurs when the state is balanced, when $\phi_\beta=\pi/2$. Error bars were computed as one standard deviation of 200 trials around each tomographic measurement, each with a random sampling of Poisson photon noise. We assume a control phase uncertainty of $\pm\pi/50$.}
\label{figure3}
\end{figure}

\subsection*{Source performance}

One of the key metrics of a photon-pair source is its pair-generation efficiency~\cite{Savanier:2016kb}. This quantity is obtained from the photon-pair detection rate as a function of the input power, accounting for loss and detector efficiency. Inside the 1-nm-wide signal and idler spectral bands, we measured a brightness of $20\, \mathrm{kHz}/ \mathrm{mW}^2$.

The indistinguishability between photon-pair sources is also important. The contrast of the RHOM block's quantum interference fringes indicates the indistinguishability of the block's constituent photon-pair sources. We measured RHOM quantum interference fringes on each source by configuring the chip to maximise photon flux at the $\mathrm{OUT}_T'$ and $\mathrm{OUT}_B'$ 
outputs, then varying $\phi_T$ and $\phi_B$ to obtain the fringes of Figure \ref{figure2}. We pumped the bottom source via the auxiliary input $\mathrm{IN}'$, and the top source via $\mathrm{IN}$ and the state-control MZI, integrating each point for 5~s. We observed $C = 93.2 \pm 1.4\%$ and $72.9 \pm 0.8\%$ fringe contrasts, respectively, for the top and bottom sources. Here, $C = (N_{\mathrm{max}}-N_{\mathrm{min}}) / (N_{\mathrm{max}} + N_{\mathrm{min}})$, where $N_{\mathrm{max}}$ and $N_{\mathrm{min}}$ are the accidental-subtracted maximum and minimum fitted count rates. The reduced contrasts can be explained by deviations (from the ideal $\eta = 50\%$) in the input evanescent couplers of each RHOM structure; they are compatible with  reflectivity values of $\eta \approx 43 \%$ and  $\eta \approx 36 \%$ for the top and bottom sources, respectively.

\subsection*{Quantum logic}

\begin{figure}[h!]
\centering
\includegraphics[width=\linewidth]{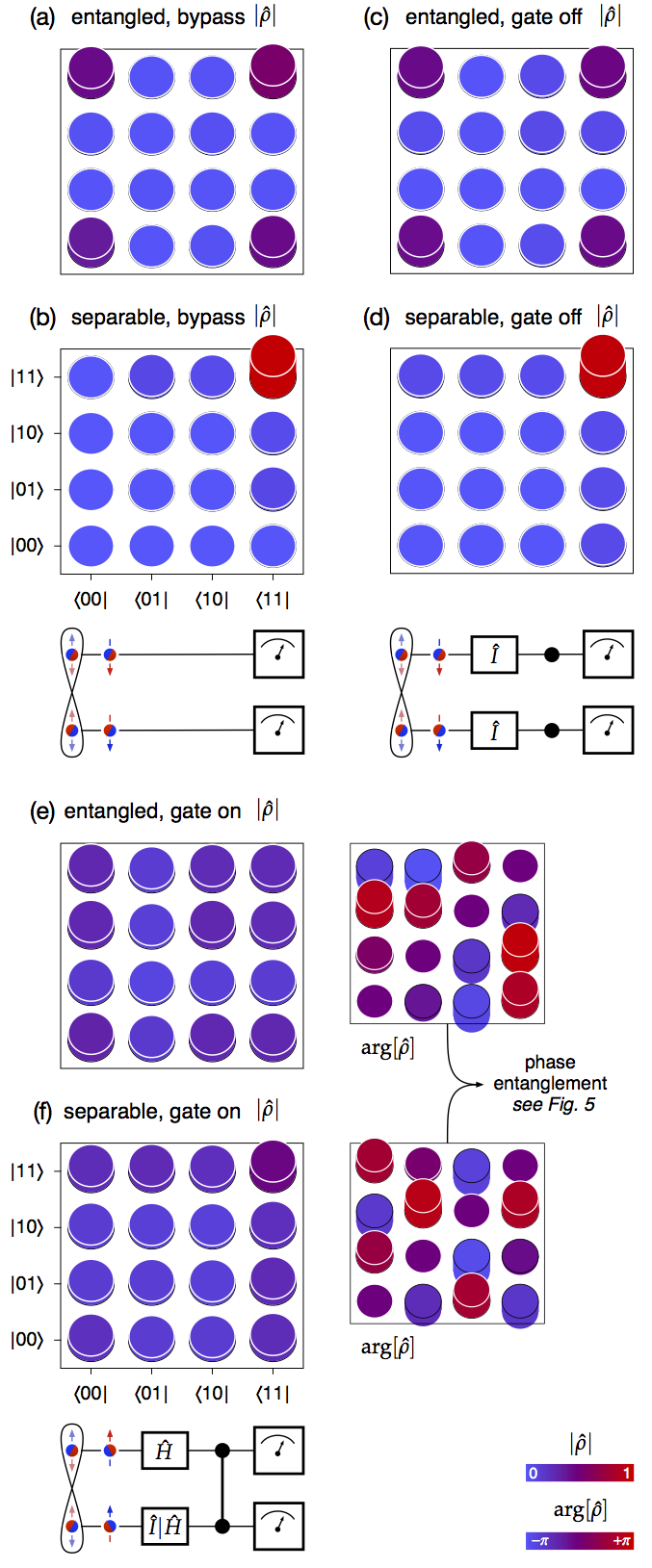}
\caption{Reconstructed output states for various source and gate configurations. States \textbf{a,c,e}~are seeded by an entangled source state, while \textbf{b,d,f}~are seeded by a $|11\rangle$ source state. States \textbf{a,b}~bypass the gate; \textbf{c,d}~pass through the gate set to $\hat{I}$; and \textbf{e,f}~pass through the gate set to $\hat{\textsc{cz}}$, and include the phase information, below. State properties are compiled in Table~\ref{table:data}. Device configurations producing each set of states are shown at right.}\label{figure4}
\end{figure}

\begin{figure}[h!]
\centering
\includegraphics[width=\linewidth]{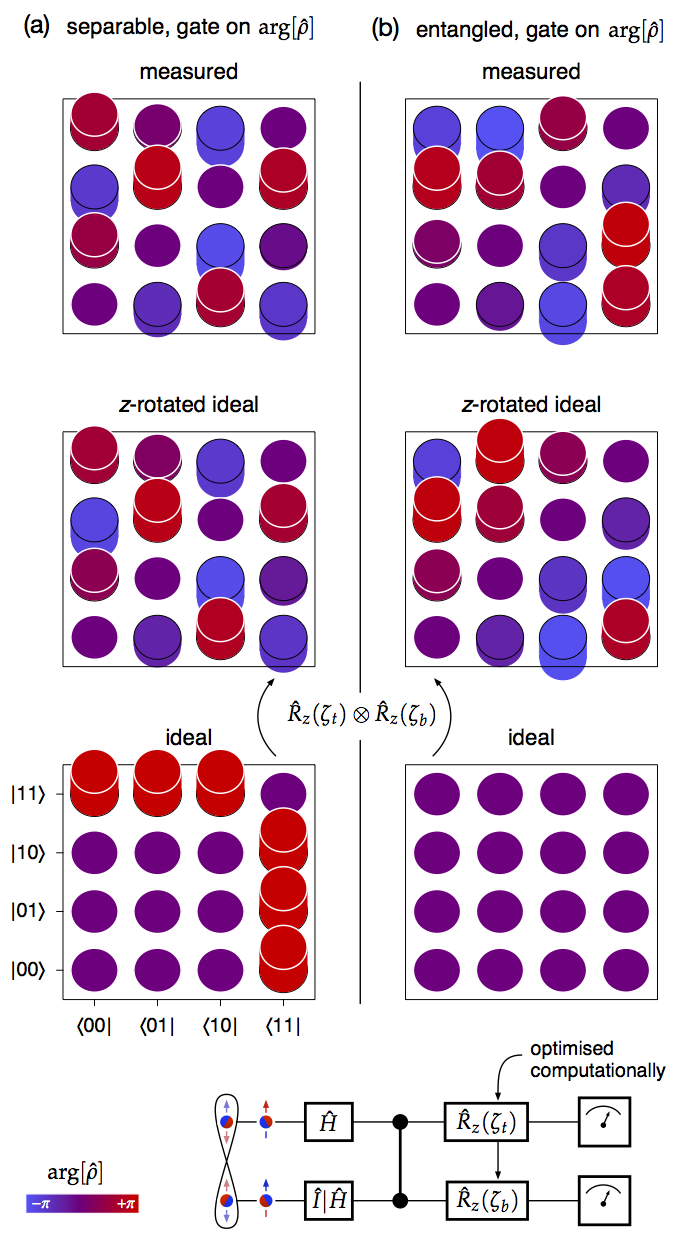}
\caption{Detail of phase entanglement, separability of states shown in Fig. Figure \ref{figure4}~e,f. Since the $\hat{\textsc{cz}}$ gate gate operates on phase, random, fixed, local $z$-rotations obscure the underlying performance. The connection between the measured and ideal states, via numerical optimisation of $\zeta_t$ and $\zeta_b$, is shown for \textbf{a} the gate-entangled, and \textbf{b} gate-disentangled states. In both cases the ideal density matrix magnitude is constant, $|\hat\rho_{i,j}| = 1/4$.}\label{figure5}
\end{figure}

We next quantified the device's control over entanglement. Quantum state tomography was used to extract the Purity ($P=\mathrm{Tr}(\hat\rho^2)$~\cite{Gamel:2012hm}), the CHSH parameter, a strict measurement of quantum correlations, and the Schmidt number, analogous to the number of pure states represented in a given density matrix. These last two metrics show how separable the state is. The CHSH inequality, $S(\hat\rho) \le 2$~\cite{Aspect:1982br, Horodecki:2009gb, Silverstone2015}, 
is violated when the state $\hat\rho$ cannot be represented by a local classical theory, indicating its entangled quantum nature. The Schmidt number, on the other hand, is an entanglement monotone and can give further evidence of the entangled or separable nature of $\hat\rho$~\cite{Terhal:2000gd, Horodecki:2009gb, Sperling:2011di}. CHSH parameter values were obtained by computationally selecting an optimal measurement set for each of the states under analysis~\cite{Silverstone2015}.

We analysed a wide set of separable and entangled quantum states produced by the two-qubit source. Fixing $\phi_T = \phi_B = \pi/2$, we varied the phase of the state control MZI, $\phi_\beta$, between $0$ and $\pi$ to prepare variably entangled states in the form of \eqref{equation1}. When $\beta = 0$ or $1$, separable states result, while  when $\beta = 1/2$, a maximally entangled state is produced. States obtained directly from the source (bypassing the gate) showed good agreement with \eqref{equation1}. These were measured using the {$\mathrm{OUT}_T'$} and {$\mathrm{OUT}_B'$} auxiliary outputs (see {Figure \ref{figure1}b}). Measured and calculated variations of the balance, Schmidt number, and CHSH parameter are plotted in Figure \ref{figure3}, versus the state control parameter $\phi_\beta$. 

In Figure \ref{figure4} we show a sample of density matrices arising from the main device configurations, and we list their properties (purity, Schmidt number, CHSH parameter, and fidelity with the ideal $z$-rotated state) in Table~\ref{table:data}. Errors were obtained from Monte-Carlo simulations, based on 200 samples of Poissonian photon noise and accompanying tomographic reconstructions~\cite{Roos:2004hm}. As expected, the $\hat{I}$-mode gate did not substantially affect the properties of the input states. The $\hat{\textsc{cz}}$-mode gate, however, acted to entangle separable states, and separate entangled states, though it also degraded the purity. The limited contrast in the quantum interference of the two RHOM sources contributed to this reduction, by occasionally depositing two photons into one `qubit'. Gate and tomography calibration errors likely also contributed. 

Since the entangling gate operates on the input state's \emph{phase}, we must examine with care the phase of the output state, $\mathrm{arg}[\hat\rho]$. The intrinsic and uncalibrated $z$-rotations on each qubit result in complicated phase pictures (Figure \ref{figure4}e,f). {To compare these to their ideal counterparts, we computationally applied $\hat R_z(\zeta_t) \otimes \hat R_z(\zeta_b)$ to the reconstructed output state, and optimised the fidelity over local $z$-rotations via $\zeta_t$ and $\zeta_b$. The resulting fidelities are listed in {Table~\ref{table:data}} and the process is shown visually in Figure \ref{figure5}.}

\section*{Discussion}

We have presented a silicon-on-insulator quantum photonic device which embeds capabilities for the generation, manipulation, and analysis of two-qubit entangled states, by leveraging on-chip linear and non-linear optics. We showed how the device can prepare a variety of entangled and separable states, and operate on them using a switchable entangling gate. We demonstrated a new reconfigurable source of variably path-entangled non-degenerate photon pairs, using reversed Hong-Ou-Mandel quantum interference, and used on-chip quantum state tomography to measure its performance. 
The integration of this source with a complex integrated linear optical network enabled both the entanglement and disentanglement of the on-chip generated quantum states.

Device performance was hindered by imperfect beam-splitters and high coupling losses, leading to issues with stability, and ultimately limiting the measurable purity and entanglement. However, the use of more advanced fibre couplers, such as those based on ultra-low loss gratings~\cite{Ding:2013hl}, together with adaptive methods, employing multiple imperfect MZIs for the realisation of a very high-quality one~\cite{Wilkes:2016ba}, can overcome these limitations, and enable high-performance, large-scale silicon photonic quantum devices in the near future.

\section*{Acknowledgements}

We thank Damien Bonneau, Jianwei Wang, and Dylan Mahler for valuable discussions and support. We are grateful to Alasdair Price for help with preliminary characterisation. We also thank the staff of the James Watt Nano-fabrication Centre in Glasgow. We acknowledge support from the European Union through the BBOI, and from the QUCHIP projects. M.G. Thompson acknowledges support from an Engineering and Physical Sciences Research Council (EPSRC, UK) Early Career Fellowship and from the European Research Council (ERC Grant Agreement number: 640079 QPE ERC-2014-ST). J.W. Silverstone acknowledges an EPSRC Doctoral Training Account, and a Natural Sciences and Engineering Research Council (Canada) Alexander Graham Bell Canada Graduate Scholarship. J.L.O'B. acknowledges a Royal Society Wolfson Merit Award and a Royal Academy of Engineering Chair in Emerging Technologies.

\section*{Author contributions statement}

R.S. and J.W.S. contributed equally to this work. They conceived and designed the device, performed the experiments, and analysed the data. M.J.S. and M. Sorel. fabricated the device. S.M., T.Y., M.F., M. Sasaki, and H.T. provided the superconducting detectors and M.G. Tanner, C.M.N., and R.H.H. built the detector system. M.G. Thompson supervised the work. All authors contributed to the manuscript.

\bibliography{biblio}

\end{document}